\documentstyle[12pt]{article}

\begin{document}
\thispagestyle{empty}
\begin{flushright}
{\large SPBU-IP-94-7}
\end{flushright}
\vspace*{2cm}

\begin{center}
{\bf \large

THE ROLE OF SOLVABLE GROUPS\\[0.5cm]
IN QUANTIZATION OF LIE ALGEBRAS}\\[1cm]

{\bf
 V.D.Lyakhovsky \footnote
 {  Supported by Russian Foundation for
 Fundamental Research, Grant N 94-01-01157-a.}
 \footnote{ E-mail address: LYAKHOVSKY @ NIIF.SPB.SU } \\
 Theoretical Department \\
 Institute of Physics \\
 St.Petersburg State University\\
 St.Petersburg \\
 198904, Russia} \\[1cm]
  (To be published in {\it Zapiski Nauchn. Semin. POMI}, V. 209) \\[1cm]
  {\bf Abstract.}
\end{center}

     The elements of the wide class of quantum universal enveloping
     algebras are prooved to be Hopf algebras $H$ with spectrum $Q(H)$
     in the category of groups. Such quantum algebras are quantum
     groups for simply connected solvable Lie groups $P(H)$. This
     provides utilities for a new algorithm of constructing quantum
     algebras especially useful for nonsemisimple ones. The
     quantization procedure can be carried out over an arbitrary field.
     The properties of the algorithm are demonstrated on examples.

\newpage
\setcounter{page}{1}

  Algebraic approach to quantization of Lie groups
\cite{Drin,Jimb} implies the following sequence of constructions
\begin{equation}
G \Longrightarrow Fun(G) \Longrightarrow U(A) \Longrightarrow
U_{q} (A) \Longrightarrow (U_{q} (A))^{*} \Longrightarrow G_{q}
\label{eq4-45}
\end{equation}
Here $U(A)$ is the universal enveloping algebra for Lie algebra $A$ of group
$ G $, $ (U_{q} (A))^* $ -- the Hopf algebra dual to
$ U_{q} (A) $.
The quantum group
$ G_{q} $ is understood as the spectrum of the
Hopf algebra $ (U_{q}(A))^{*} $, i.e.
it corresponds to
$ (U_{q}(A))^{*} $
just as the group $ G $ to the algebra of functions $ Fun(G) $.
In the sequence (\ref{eq4-45}) the main procedure is the construction
of the quantum algebra $ U(A) \Longrightarrow U_{q}(A) $.
  It was stressed in \cite{Fadd} that the main
object in the quantization procedure is the quantum algebra of functions
$ Fun_q(G) \approx (U_q(A))^* $, moreover this interpretation is also valid
for $U_q(A)$ (but with respect to the dual group). At the same
time in general situation the spectrum of a noncommutative Hopf algebra
does not exist and the pair
$ Fun_q(G) \Longrightarrow G_q $ must be understood
as unique entity.

  The problem of equivalence between the categories $U_q(A)$
and $ (U_{q}(A))^{*} $ was first mentioned in \cite{Drin}, where
the enveloping algebra $U_q(A)$ was incorporated in the family of Hopf
algebras having the commutative classical limit.
The quantum duality principle
 \cite{Seme} allows to elucidate the nature of this equivalence.
If one consideres the Lie bialgebra $ (A,A^*) $
as the starting object \cite{Drin} then the quantization is
the simultaneous deformation of
$ Fun(G) $ and $ Fun(G^*) $, where $ G $ , $ G^* $
are the Lie groups with algebras $ A $  and $ A^* $.
So $ Fun_{q} (G^*) $ (as an algebra dual to
 $ Fun_q (G) \approx (U_q (A))^{*} $ ) is equivalent to the Hopf algebra
$ U_q (A) $ thus being not only a quantum algebra but also a
quantum group.

  In this paper the scheme is proposed for the explicit
realization of $U_q(A)$ as a quantum group. It is shown that the wide class
of quantum universal enveloping algebras are factor algebras of such
noncommutative Hopf algebras $H$ that the spectrum $Q(H)$ does exist.
The 'classical limit' of $Q(H)$ is a simply connected solvable group
$P(H)$ -- the factor group of $G^*$. The established correspondence between
$U_q(A)$ and solvable Lie groups $P(H)$ provides new possibilities
to construct quantum Lie algebras.

  In sect.1 the properties of the selected class of Hopf algebras
$U_q(A)$ are formulated. For each $U_q(A)$ the simply connected
solvable Lie group $P(H)$ strictly corresponds.
In sect.2 the inverse problem is solved -- for a given group $P$ the
Hopf algebra $H$ is obtained that can be factorized to the quantum
universal enveloping algebra $U_q(A)$.
In sect.3 the exposed scheme is demon\-strated on some known
constructions and an example of application of the new quantization
method is given.

  For the elements of tensor products of Hopf algebras
the abbreviated notation will be used:
$ a' \equiv a \otimes {\bf 1}; \rule{0.6cm}{0cm}
a'' \equiv {\bf 1} \otimes a. $       \\

{\large \bf \underline{1.} }

   Consider the quantum universal enveloping algebra
$ U_{q} (A) $ of an algebra $A$ over an arbitrary field
 $ {\bf K} $ with generators
$ \{ x_{l} \}, \;\; l = 1, \ldots, n $ and a set of quantization
parameters $q$.
Set the following conditions

u.1) for generators $ \{ x_{l} \} $ the tensor multipliers
$\Delta'_j$ and $\Delta''_j$ in the coproduct
\[ \Delta (x_l) = \Sigma_j \Delta'_j(x_l) \Delta''_j(x_l) \]
are either linear functions of generators, or convergent power series
in $\{ h_i \}$. The elements of
$ \{ h_i \} , \rule{0.5cm}{0cm} i=1,\ldots,m<n \rule{0.5cm}{0cm} $, commute.

u.2) in the coproduct
\[ \Delta (h_i) = \Sigma_j \Delta'_j(h_i) \Delta''_j(h_i) \]
$\Delta'$ and $\Delta''$ transfer $h_i$ to the subalgebra $U_q^{(h)}$
generated by $\{ h_i, {\bf 1} \}$.

u.3) $ \varepsilon (x_k) = 0. $

u.4) the relations
$ ( \cdot )(S \otimes \mbox{id}) \Delta =
( \cdot )( \mbox{id} \otimes S) \Delta = \eta \circ \varepsilon $,
applied to generators $\{ x_l \}$, can be solved (when subalgebra $U_q^{(h)}$
is commutative) to define all the $S(x_l)$,
whatever the other multiplications in  $U_q(A)$ are.

u.5) $ \lim_{q \rightarrow 0} \Delta(x_l) = x_l' + x_l''. $

  It is known \cite{Abe} that for each pair $ (X,R) $ of Hopf
algebra $X$ and
commutative unital algebra  $ R $ the coproduct
$ \Delta_{X} $, counit $ \varepsilon_{X} $ and antipode
$ S_{X} $
induce the group structure on the set of algebraic
morphisms $ {\rm Hom}(X,R) $
with the multiplication
\begin{equation}
( \chi_1 * \chi_2 ) (x) = (\cdot)_{R} (\chi_1 \otimes \chi_2)
\Delta_{\chi} (x);  \rule{0.5cm}{0cm}
x \in X; \;\;\; \chi_1,\chi_2 \in {\rm Hom}(X,R).
\label{eq4-46}
\end{equation}

When $ R $ is noncommutative the previous statement fails, the map
$ \chi \circ S_{X} $ having no more the property of inverse
element.
The fact is that $ S_{X} $ being an antihomomorphism  forms the
composition $ \chi \circ S_{X}$ that does not belong to
$ {\rm Hom}(X,R) $ .

  We shall demonstrate that for algebras $U_q(A)$
(with the properties  u.1 - u.5) this obstacle can be overcome.
Let us consider $U_q(A)$ together with such an associative algebra
$H$ (with the same set of generators) that

h.1) the operators $ \Delta_H, S_H, \varepsilon_H $
and $ \eta_H $ on the generators coinside with the corresponding
defining compositions in $U_q(A)$,

h.2) the subalgebra $H^{(h)}$,
generated by $ \{ h_i, {\bf 1} \} $,
is equivalent to $U_q^{(h)}$,

h.3) the algebra $H$ is
free modulo the relations of commutativity of $H^{(h)}$ ,

h.4) the operators $\Delta$ and $\varepsilon$ are extended to $H$
homomorphically and the antipode $S$ -- antihomomorphically.

  These properties guarantee that $H$ is a Hopf algebra.

  Let $V$ and $V^{(h)}$ be the subspaces of the vector space of
 $H$ -- the correspond\-ing lineals of $\{ x_l \}$ and
$\{ h_i \}$ .
Consider a free associative algebra $L$  and the space of moprphisms
Mor$(V,L)$. In Mor$(V,L)$ define the subset
 $ \mbox{Mor}^{(h)} $ such that its elements send the space
$(V^{(h)})$ to the fixed commutative subalgebra in $L$.
The set
$ \mbox{Mor}^{(h)} $ is obviously a vector space. Each
$ \zeta \in  \mbox{Mor}^{(h)} $ is fixed by
$n$ coordinates $ \zeta (x_l)$. Let $\zeta_{\uparrow H}$
be the homomorphic extension of $\zeta$ to $H$. Such
extensions always exist but do not constitute the vector space anymore.

  The multiplication on $ \mbox{Mor}^{(h)} $
will be introduced similarly to (\ref{eq4-46}):
\begin{equation}
\zeta_1 * \zeta_2 = (\cdot )_L (\zeta_{1 \uparrow H} \otimes
\zeta_{2 \uparrow H}) \Delta .
 \label{eq4-48}
 \end{equation}
For each $ \zeta \in \mbox{Mor}^{(h)} $ the inverse will be given by
\begin{equation}
\zeta^{-1} \equiv \zeta_{\uparrow H} \circ S.
 \label{eq4-49}
 \end{equation}
Note that according to the definition of $ \zeta $ the antipode $S$
in $\zeta^{-1}$ acts only on the linear combinations of the generators.
{}From  u.2 and u.4 it follows that
 $ \zeta^{-1} \in \mbox{Mor}^{(h)} $ for any
$ \zeta \in \mbox{Mor}^{(h)} $. The map
\begin{equation}
\zeta_{(0)} \equiv \eta_L \circ \varepsilon_H
  \label{eq4-51}
  \end{equation}
is the zero vector in the space
 $ \mbox{Mor}^{(h)} $ (see the property u.3).

  Let $G$ be the Lie group with the algebra $A$. Denote by
$Q(H)$ the space $ \mbox{Mor}^{(h)} $
with the multiplication (\ref{eq4-48}),
the inversion (\ref{eq4-49}) and the marked element (\ref{eq4-51}).

 \underline{{\large Proposition 1.}}
  $Q(H)$ is a group. The 1-dimensional
representation $d$ of $L$ transforms the group
$Q(H)$ into the vector solvable Lie group $P(H)$
on the $n$-dimensional vector space. Groups $P(H)$ and
$G^*$ are equivalent if and only if ${\rm dim}G^*=n$.

 \underline{{\large Proof.}}
  Consider the product
\[ \begin{array}{ccc}
\zeta * \zeta^{-1} = &
(\cdot )_L ( \zeta_{\uparrow H} \otimes \zeta_{\uparrow H} \circ
S_{\uparrow H}) \Delta = & \\
  & = (\cdot )_L (\zeta_{\uparrow H} \otimes \zeta_{\uparrow H} )
  ( \mbox{id} \otimes S_{\uparrow H} ) \Delta = &
  \zeta_{\uparrow H} (\cdot )_H ( \mbox{id} \otimes S_{\uparrow H} )
  \Delta.
  \end{array}
  \]
Note that $S_{\uparrow H}$, used here according to the definition
 (\ref{eq4-48}), is not the antipode of
$H$. This operator coinsides with $S_H$ on generators and is
 homomorphically extended to $H$. Nevertheless the properties
 u.1  and  u.2 guarantee that the operator $S_{\uparrow H}$
 in the multiplication of $Q(H)$
 acts either on  $V$, or on power series in
 $U^{(h)}$. In these situations $S_{\uparrow H}$
 coinsides with $S$ and the last equality can be continued:
\begin{equation}
\begin{array}{cc}
\zeta * \zeta^{-1} = &
  \zeta_{\uparrow H} (\cdot )_H ( \mbox{id} \otimes S ) \Delta = \\
  \zeta_{\uparrow H} \eta_H \varepsilon_H = &
 \eta_L \varepsilon_H = \zeta_{(0)}
 \end{array}
\label{eq4-52}
  \end{equation}
 The properties of $ \zeta_{(0)} $ (as the unit of $Q(H)$) and
 the associativity of multipli\-ca\-tion are verified similarly.

  The representation $ d: L \rightarrow R $ maps $L$
in the abelian algebra $R$ and induces the transformation of the space
  $ \mbox{Mor}^{(h)} $ to
$\mbox{Mor}(V,R)$. The elements
 \[ d \circ \zeta \in \mbox{Mor}(V,R) \]
are fixed by the coordinates
 $ \{ ( d \circ \zeta )(x_k) \} $. Given the properties  u.1 - u.3
 the coordinates of the product are the analitical functions of the
 coordinates of factors. Thus $P(H)$
 is the  $n$-dimensional vector space with an analitic group
 multiplication law. Due to Levy theorem the vector Lie group is a
 solvable group with trivial tori.

 \underline{{\large Corollary}}
  The group $P(H)$ can be presented as a sequence of semidirect
 products of vector spaces (as abelian additive groups).

  The obtained sequence
\begin{equation}
U_q(A) \Rightarrow H \Rightarrow Q(H) \Rightarrow P(H)
\label{eq4-53}
\end{equation}
 is unique and can serve for classification of quantum algebras
 $U_q(A)$. The group $Q(H)$ in (\ref{eq4-53}) is the group with
 noncommuting coordinates. It is the quantum analogue of the group
 $P(H)$. The properties of the coordinate algebra
 $L$ are defined by the multiplication in  $H$.
 Abelian subalgebra in $L$ is fixed by the images
  $\zeta(H^{(h)})$, all other
   multiplications in $L$ are free.
 For the algebra $H$ the existence of groups
  $Q(H)$ and $P(H)$ means that $H$ retains the properties of Hopf algebra
 when its multiplication is changed for abelian one while the coalgebra
 structure is conserved. Such algebra $H^c$
 is evidently an algebra of functions on the dual group $P(H)$.
 $H$ itself is an algebra of noncommutative functions on $P(H)$.
 The Hopf algebra $U_q(A)$ can be treated as
 $ \mbox{Fun}_q(G^*)$ , as $ \mbox{Fun}_q(P(H))$ and also as
 $ \mbox{Fun}_q(Q(H))$. To realize the last variant one must consider
 the quantum commutation relations as a deformation of almost free
 multiplication (strictly speaking we have here the coboundary
 deformation, i.e. the contraction \cite{Lya1}).
\\

{\large \bf \underline{2.} }

  Let us turn to the problem of construction of $U_q(A)$-type
 algebra starting with the solvable Lie group. Fix the global
 coordinate system on the simply connected solvable Lie group $P$ .
 Let $Q$ be the solvable group with coordinates in an
 associative algebra $L$. Suppose that after the change of scalars
 $L \rightarrow L/L^{ (1) } $ the group $Q$ becomes equivalent to
 $P$. (Here $L^{ (1) }$ is the first derivative of $L$.)
 The multiplication on $L$ can have additional restrictions
 when the standard group axioms on $Q$ are imposed .
   Consider for example
\begin{equation}
Q = \mbox{\bf R}^p \triangleright \mbox{\bf R}^q,
\label{eq4-54}
\end{equation}
where the subgroups $ \mbox{\bf R}^p$ and $\mbox{\bf R}^q$
are the abelian additive and the multiplication is goverened by
the homomorphism $ \Phi : \mbox{\bf R}^p \rightarrow \mbox{Aut}
(\mbox{\bf R}^q) $,
\[
(a',b')(a'',b'') = (a'+a'', \Phi (a'') b' + b'')
\]
Suppose the coordinates $\{ a_s, b_t \}$ belong to $L$. Then
the associativity and the inverse element properties imply the
commutativity of coordinates $\{ a_s \}$. We shall not discuss here
the conditions under which the noncommuta\-tive algebra $L$ exists realizing
the transformation of $P$ into $Q$. Suppose $L$ is such an algebra
(with the necessary multiplication properties).
Suppose also that the coordinate system of $Q$
is correlated with the analitic coordinates on $P$.
   Construct the coalgebra $H$, generated by the coordinate functions
 $\Psi_i$ on $Q$:
 \begin{equation}
 (\Delta \Psi_i ) (\phi' \otimes \phi'') = \Psi_i (\phi' \cdot \phi''),
 \label{eq4-56}
 \end{equation}
 \begin{equation}
 \varepsilon (\Psi_i) = \Psi_i (e_Q) = 0, \rule{0.5cm}{0cm}
 \eta \,(1) = {\bf 1}_H,
 \label{eq4-57}
 \end{equation}
 where $\phi',\phi'' \in Q$. The number of generators in $H$
 is equal to dim$P$. Define the multiplication in $H$ ,
 \begin{equation}
 ( \Psi_i \cdot \Psi_j ) (\phi) = \Psi_i (\phi) \Psi_j (\phi)
 \label{eq4-58} \end{equation}
and the antipode
 \begin{equation}
 (S  \Psi_i ) (\phi) = \Psi_i (\phi^{-1}).
 \label{eq4-59} \end{equation}
The direct verification of Hopf algebra axioms proves the validity
 of the following statement.

 {\large \underline{Proposition 2.}}
   $H$ is a Hopf algebra if the operations
$\Delta$ and $\varepsilon$ are extended to $H$ by homomorphisms and
 $S$ -- by antihomomorphisms.

 {\large \underline{Note.}} The properties
  (\ref{eq4-56})  and (\ref{eq4-57}) are true for an arbitrary element of
 $ H$, while the antipode of a policoordinate function is not equal to
 the function of inverse element.

  We have constructed such a noncommutative and (in general)
noncocom- \\ mutative Hopf algebra $H$,
that $Q$ is its quantum group.
 Only on the subset of generators of $H$ the multiplication
 defined by (\ref{eq4-58}) is free.

  We are interested in the ideals of $H$ which guarantee  that
 $H/J(H)=U_q(A)$.
 The structure of $J(H)$ must be in agreement with operations
  (\ref{eq4-56},  \ref{eq4-57},\ref{eq4-59})
 and the multiplication (\ref{eq4-58}).
 \begin{equation}
 [ \Psi_i , \Psi_j ] = \Phi_{ij}, \rule{1cm}{0cm} \Phi_{ij} \in H
 \label{eq4-60}
 \end{equation}
 The Jacoby identity and the correlation conditions
 \begin{equation}
 \Delta \Phi_{ij} = [\Delta \Psi_i ,\Delta \Psi_j ], \rule{0.5cm}{0cm}
 S \Phi_{ij} = -[S \Psi_i ,S \Psi_j ]
 \label{eq4-61}
 \end{equation}
 give the system of equations for the deforming functions $ \Phi_{ij}
 (\Psi_l)$. Every solution of this system provides
 $H$ the properties of quantum universal enve\-loping algebra
 $U_q$.

  To find the algebra  $U(A)$
 corresponding to the classical limit of $U_q$ let us contract
 the group $P$ to the Abelian vector group.
 This means that  $P$ is included in the
 $q$-parametric set of deformations of Abelian group lim$(P_{q})$
 treated as additive.
 For the groups $Q_q$ and the algebras
 $H_q$ and $U_q$ the same parametrisation is induced. If the limit
 lim$U_q \equiv U(A)$ exists the algebra
 $U_q$ (the quantum analogue of the group
 $P$ ) can also be considered as a quantum universal enveloping algebra of
 $A$. Note that the dimension of $A$ may be infinite.

  Such an approach treats the quantization of a Lie algebra $A$
 as a deformation of an Abelian vector group to a solvable one.

  It must be pointed out that in this approach one can
 pay no attention
 to simplicity or semisimplicity of  $A$, as well as to the properties
 of its main field.
 Thus one gets the possibility to construct quantum analogues
 of nonsemi\-simple Lie algebras and to obtain directly the real quantum
 algebras.  \\

{\large \bf \underline{3.}}

{\large \bf \underline{3.1}}
  For the standard quantization of simple complex Lie algebras
 \cite{Drin,Jimb} the conditions u.1 - u.5 are valid.
 Here the solvable group $P(H)$
 can be treated as a group of upper triangular matrices
\begin{equation}
\left(
\begin{array}{cccccc}
e^{\textstyle 1/2(q_1 h_1)} & 0 & \cdots & 0
& 0 & e^{\textstyle 1/4(q_1 h_1)}x_1^+ \\
0 & e^{\textstyle 1/2(q_1 h_1)} & \cdots & 0
& 0 & e^{\textstyle 1/4(q_1 h_1)}x_1^- \\
\cdots & \cdots & \cdots & \cdots
& \cdots & \cdots \\ 0 & 0 & \cdots
& e^{\textstyle 1/2(q_r h_r)} & 0 &
e^{\textstyle 1/4(q_r h_r)}x_r^+ \\ 0 & 0 & \cdots  & 0
& e^{\textstyle 1/2(q_r h_r)} & e^{\textstyle 1/4(q_r h_r)}x_r^- \\ 0 & 0
& \cdots  & 0 & 0 &        1
\end{array} \right) \label{eq4-62}
\end{equation}
 with coordinates $\{ h_i, x_i^{\textstyle \pm} \} \,\,$,
 and $r$ equal to the rank of
 $A$. Let $L_0$ be a free associative algebra with generators
 $\{ h_i, x_i^{\textstyle \pm}, {\bf 1} \}$. Factorize it by the ideal,
 generated by the commutators of $\{ h_i \}$. The group structure is
 preserved if the coordinates of the matrix group
(\ref{eq4-62}) belong to the factor algebra of $L$.
 We get the group  $Q(H)$. The auxiliary Hopf algebra $H$ is a factor
 of the free algebra generated by
 coordinate functions  $\{ H_i, X_i^{\textstyle \pm },
{\bf 1} \}$ on $Q(H)$ with operations
$ (\cdot), \Delta,S,\varepsilon$ and $\eta$ given
 by (\ref{eq4-56} - \ref{eq4-59}). The ideal $J(H)$ is generated by
 the relations $H_iH_j = H_jH_i$. The final result --
 the quantum algebra $U_q(A)$ -- can be obtained
 when the correlation equations are solved
 and the deformations of  $U(A)$ are in agreement with the operations
(\ref{eq4-56},\ref{eq4-57}) and (\ref{eq4-59}). Algebra
$U_q(A)$ is thus realized as a quantum group $\mbox{Fun}_q(P(H))$. In the
only case when the number of generators in $U_q(A)$ is equal to the dimension
of $A$, the group
$P(H)$ is equivalent to the dual group $G^*$.

{\large \bf \underline{3.2} }
  In \cite{Lya2} the solvable groups of the type (\ref{eq4-54})
were used to construct quantum algebras $U_q(A)$.
In the matrix form the vector group $P(H)$ that gives the Hopf algebra $H$
can be written as
\begin{equation}
\left(
\begin{array}{cc}
e^{\textstyle \gamma^ih_i} & e^{\textstyle -\beta^ih_i }x \\
 \cdots &  \cdots \\
0  \cdots   0 &        1
\end{array} \right) \label{eq4-63}
\end{equation}
Here $x$ is an $m$-dimensional vector,
$ \{ \gamma^i, \beta^i \}$ is the set of commuting
 $ m \times m $-matrices, $ i=1, \dots, u; u \leq m $.
Parameters $\{ h_i \}$ are the coordinates of the abelian matrix group
$ e^{(\textstyle \gamma^ih_i)}       $.
 Let the matrices (\ref{eq4-63}) have the noncommutative coordinates
 belonging to the associative algebra $L$ with generators
 $ \{ h_i,x_i, {\bf 1} \}$ and relations $h_ih_j = h_jh_i$.
 The coordinate functions $\{H_i,X_j,{\bf 1 } \}$
 generate algebra $H$. Its multiplication properties are analogous to
 those of $L$ while the coproduct has the form
\[
(\Delta X)_j = (e^{\textstyle \alpha^i H_i'} X''
+ e^{\textstyle \beta^i H_i''} X')_j,
\rule{0.3cm}{0cm}
\Delta H_i = H_i' + H_i'',
\rule{0.3cm}{0cm}
\alpha^i \equiv  \beta^i + \gamma^i .
\]
It was shown in \cite{Lya2} that in vertue of such properties $H$ is a
Hopf algebra. It follows from the Proposition 1 that the group
stucture is conserved when the numeric coordinates in $P(H)$ are
substituted by the noncommutative ones. In other words there exists
the group $Q(H)$ that is the spectrum of the algebra $H$.
With the help of $Q(H)$ and $H$ the quantum analogues of
nonsemisimple real algebras where constructed in \cite{Lya2} for
Heisenberg and 2-dimensional flat motions algebra.

{\large \underline{3.3}}
 Let $P$ be the nilpotent matrix group:
 \begin{equation}
 \left( \begin{array}{ccc}
 1 & x_1 & y \\
 0 & 1   & x_2 \\
 0 & 0   & 1       \end{array} \right)
       \label{eq4-79}
        \end{equation}
its coordinates $ x_i,y $ can belong to an arbitrary associative algebra.
Consider
$L$ to be freely generated by $ \{ x_i, y, {\bf 1} \} $ .
The group $Q$ is thus defined. The free associative algebra
$H$ with the generators $ \{ X_i, Y, {\bf 1} \} $
(dual to $ \{ x_i, y, {\bf 1} \} $ )
will be supplied by the coproduct
\begin{equation}
\Delta X_i = X_i' + X_i'', \rule{1cm}{0cm} \Delta Y = Y' + Y'' + X_1'X_2'',
        \label{eq4-80}
        \end{equation}
the antipode
\begin{equation}
S X_i = - X_i, \rule{1cm}{0cm} S Y = - Y  + X_1 X_2,
        \label{eq4-81}
        \end{equation}
and the counit
\begin{equation}
\varepsilon  (X_i) = \varepsilon ( Y ) = 0.
        \label{eq4-82}
        \end{equation}
It is evident that the commutative Hopf algebra $ H^c$
with the properties (\ref{eq4-80} - \ref{eq4-82})
also exists.
Thus we shall search for a deformation $U_q$  of
 $(H^c)$:
\[ [ X_i, Y ] = \Phi_i, \rule{1cm}{0cm} [ X_1, X_2 ] = \Phi_{12}. \]
Equations (\ref{eq4-61}) are easily solved. The verification
of the Jacoby identities completes the construction of $U_q$.
It is defined by formulas (\ref{eq4-80} - \ref{eq4-82}) and the Lie
composition
\[ [ X_1, X_2 ] = X_1 + X_2, \rule{0.5cm}{0cm}
   [ X_1, Y ] =  Y + (1/2) X_1  X_1, \rule{0.5cm}{0cm}
   [ X_2, Y ] = -Y - (1/2) X_2  X_2, \]

One can easily obtain the contineous family of groups $P(H;q)$
with the coordinates $ \{ x_i, y \} $ and the multiplication
\[ ( x_i', y')(x_i'', y'') = ( x_i' + x_i'', y' + y'' + qx_1' x_2'' ).
\]
This leads to the corresponding family of algebras $U_q$:
\[
\begin{array}{ll}
\Delta X_i = X_i' + X_I'', & [ X_1, X_2 ] = X_1 + X_2, \\
\Delta Y   = Y' + Y'' + q X_1' X_2'', & [ X_1, Y ] =  Y + (q/2) X_1  X_1,
\\
S (X_i) = - X_i, & [ X_2, Y ] = -Y - (q/2) X_2  X_2. \\
S (Y)  = - Y + qX_1X_2,
& \varepsilon (X_i) = \varepsilon (Y) = 0.
\end{array}
\]
In the limit $ q \rightarrow 0 $ the family $U_q$ tends to the universal
enveloping algebra $U(A)$, where
$A$ is defined by the compositions
\[
[ X_1, X_1 + X_2 ] = X_1 + X_2, \rule{0.5cm}{0cm}
[ X_1, Y ] =  Y, \rule{0.5cm}{0cm}
[ X_1 + X_2, Y ] =  0.
\]
This proves the Hopf algebra
 $U_q$ to be the quantization of the Lie algebra
$A$. At the same time  $U_q$ is realised as the quantum group for
the group $P$: $ U_q \approx \mbox{Fun}_q(P)$.
Let $G$ be the simply connected Lie group with algebra $A$.
The Lie algebra $A^*$ of the group  $P$ forms together with
 $A$ the Lie bialgebra. Thus the group  $P$ is equivalent to the group
 $G^*$ dual to $G$.

  In the general case quantum universal enveloping algebra
$U_q$ of the type described here can be a quantum group
 $Fun_q$ not only for the dual group $G^*$ but also for its factor group $P$.
 The group $P$ is the minimal factor group in $G^*$ that totally defines
 the algebra $U_q$.

  The author is grateful to P.P.Kulish for
numerous stimulating discussions.

\end{document}